\documentstyle[aps]{revtex}

\begin{document}
\title{Huygens' principle, the free  Schr\"odinger
particle and the
quantum anti--centrifugal force}
\author{M. A. Cirone $^\dagger$, J. P. Dahl$^\dagger$ $^\ddagger$,
M. Fedorov$^\dagger$ $^\S$, D. Greenberger$^\P$,\\
W. P. Schleich$^\dagger$}
\address{$^\dagger$
Abteilung f\"ur Quantenphysik,
Universit\"at Ulm, D-89069 Ulm, Germany \\
$\ddagger$ Chemical Physics, Department of Chemistry,
Technical University of Denmark, DTU 207, DK-2800
Lyngby, Denmark \\
$^\S$ General Physics Institute, Russian Academy
of Sciences, 38 Vavilov st., Moscow 117942 Russia \\
$^\P$ City College of New York, New York, NY 10031, USA}

\maketitle

\begin{abstract}
Huygens' principle following from the d'Alembert wave equation
is not valid in two--dimensional space. A Schr\"odinger
particle of vanishing angular momentum
moving freely in two dimensions experiences an
attractive force -- the quantum anti--centrifugal force --
towards its center. We connect these two phenomena by
comparing and contrasting the radial propagators of the
d'Alembert wave equation and of a free
non--relativistic quantum mechanical particle in two and
three dimensions.
\end{abstract}

\section{Introduction}

In the year 1917 Paul Ehrenfest \cite{ehr} addressed
the question: Why is the space we live in
three--dimensional? He provided three physical phenomena
known at his time that crucially depend on the number
of space dimensions: (i) The {\em generalized} electrostatic
attraction leads only in two and three dimensions to stable
circular orbits,
(ii) only in three dimensions do we find as many electric as
magnetic field components and (iii) Huygens' principle,
based on the d'Alembert wave equation, is
only valid in a world with an odd number of spatial dimensions.

In the present paper we connect this phenomenon
of the breakdown of Huygens' principle in two dimensions
to the surprising appearance of an attractive
force acting on a free particle of vanishing angular
momentum in two dimensions \cite{us1}. We show
that both phenomena have the same physical
origin: Interference of waves.

We compare and contrast the Green's functions of
the free Schr\"odinger equation corresponding to
vanishing angular momentum
in two and three dimensions.
Since the rotational symmetry of the initial wave function
is preserved in the time evolution, we can confine ourselves
to the Green's function $g^{(2)}$ and $g^{(3)}$ of the
radial motion. We derive analytical expressions for
these quantities by taking the product of the
Green's functions of the individual one--dimensional
motion along each Cartesian axis and integrating over
the angular part.

We show that the radial Green's function in three dimensions is
the difference of two Green's functions
of two free one--dimensional motions: One Green's function describes
the free motion from the initial radius $\rho$
to the radius $r$. The other Green's function corresponds to
the motion of the particle starting at the mirror image
$-\rho$ of the initial radius $\rho$ reaching $r$.
In the three--dimensional space
the phase difference between these two Green's
functions is $\pi$.

In contrast, the corresponding Green's function
of the two-dimensional
radial motion is more complicated. It is again the difference
of two one-dimensional Green's functions. However,
they do not correspond to free motion anymore.
Moreover, the phase difference between the two contributions
is now $\pi/2$. This feature reflects the fact
that in two dimensions
the radial Schr\"odinger equation for vanishing angular momentum
enjoys an additional potential that is attractive and of quantum
origin. It corresponds to the quantum anti-centrifugal force.

Whereas in one and three dimensions this potential is absent,
it reappears in four and higher dimensions. However, it is repulsive.
In contrast to the breakdown of Huygens' principle,
which takes place for all even dimensions, this additional
potential displays the unique behaviour of being attractive
only in two dimensions.

Our article is organized as follows: In section 2 we
briefly review the physical origin of the breakdown of Huygens'
principle and summarize the essential features of the quantum
anti--centrifugal potential. Here we confine our discussion to
two--and three--dimensional situations.
In section 3 we derive analytical expressions for the radial
Green's function of the free Schr\"odinger equation in
two and three dimensions. In the appendix we make contact with
the corresponding expressions for the d'Alembert wave equation.

\section{Two dimensions are very different from three}
\label{two}

In the present section we review two phenomena which appear to be
unrelated at first sight-- the breakdown of Huygens' principle applied to
d'Alembert waves in two dimensions and the
quantum anti--centrifugal force acting on a Schr\"odinger particle
moving freely
in two dimensions. Here we do not dwell on mathematical
formalism but focus on the key ideas. For the sake
of completeness we include detailed derivations
of the propagators of the d'Alembert wave equation
in the appendix.

\subsection{Breakdown of Huygens' principle}
\label{bhp}

In \cite{ehr} Ehrenfest summarized earlier work by J Hadamard on the
propagation of a signal according to the d'Alembert wave equation.
When we express his findings in today's language, he investigated
the Green's function of the d'Alembert equation: A delta function
perturbation orginating at time $t=0$ from the origin travels in three
dimensions as a spherical wave which at time $t$ had reached $r$.
In this case the
perturbation always keeps its shape, since for $t>0$ the propagator
in three dimensions reads

\begin{displaymath}
G^{(3)}_{\rm ret}\left( t,r \right)
=\frac{\delta (ct-r)}{4\pi r}
\end{displaymath}
However, in two dimensions the delta function perturbation does not
propagate as a well localized
circular wave, but contains a
tail, a so--called wake field \cite{cou}.
As shown in the appendix,
the propagator takes the form

\begin{displaymath}
G^{(2)}_{\rm ret}\left( t,r \right)
=\frac{\Theta [(ct)^2-r^2]}{\sqrt{(ct)^2-r^2}}.
\end{displaymath}
Due to the Heaviside step function there exists a sharp wave front.
Moreover, together with the square root in the denominator
it creates a long tail of
excitation in the inside domain of the propagating
circular ring \footnote{In his novel {\em Flatland}
E Abbott \cite{abb} describes the phenomenon that
people living in a two--dimensional world have a dim
vision because of fog. They see unsharp
surroundings. We do not know if the author was
aware of the mathematical and physical details of  Green's
functions in two dimensions.}.
This wake field can be observed with
water waves after a stone has excited
a wave in an initially quiet pond.

It is interesting to trace the mathematical
reason for the occurrence of the square root
in the propagator back to the
dimensionality of space. Indeed, we show
in the appendix that the square root stems from the
fact that the two--dimensional volume element
reads $dr\: r\: d\varphi$. In contrast the three--dimensional
volume element $dr \:  r^2 d\theta
\sin \theta\; d\varphi$ has in the angle variable
an additional factor $\sin \theta$. It is because of this
extra factor that the interference of plane waves
in three dimensions gives rise to a spherical
Bessel function

\begin{equation}
\frac{1}{2}\int_{0}^{\pi}\!\!\! d\theta \sin \theta \:
e^{-ikr\cos \theta}=\frac{\sin kr}{kr} \equiv j_{0}(kr)
\label{sbf}
\end{equation}
whereas in two dimensions we find an ordinary Bessel function

\begin{equation}
\frac{1}{2\pi}\int_{-\pi}^{\pi}\!\!\! d\varphi \:
e^{-ikr\cos \varphi}\equiv J_{0}(kr)
\label{somm}
\end{equation}
The spherical Bessel function

\begin{equation}
j_{0}(kr)=\frac{1}{2ikr}\left( e^{ikr}-e^{-ikr}\right)
\label{spher}
\end{equation}
originating from the
interference in three dimensions is an exact superposition
of an incoming and an outgoing spherical wave
with a fixed phase difference of $\pi$.

In contrast, the familiar asymptotic expansion
\cite{grad} of the ordinary Bessel function

\begin{equation}
J_{0}(kr) \cong \sqrt{\frac{2}{\pi kr}}\cos \left(
kr-\pi/4 \right)
= \sqrt{\frac{1}{2\pi kr}}e^{-i\pi/4} \left(
e^{ikr}-ie^{-ikr}\right)
\label{appr}
\end{equation}
suggests that the interference of plane waves in two
dimensions gives rise to an approximate superposition of an
incoming and an outgoing cylindrical wave with an asymptotic
phase difference of $\pi/2$. However, equation (\ref{appr})
is only the most elementary asymptotic approximation of the
Bessel function. In the literature there exists a
variety of techniques \cite{berry} to find more accurate
expansions. Independent of their specific form, they
clearly indicate that the result of the interference of
plane waves in two dimensions is not a pure circular wave.
This also manifests itself in the fact that the zeros
of $J_{0}$ are not equally distributed like $j_{0}$
but rather bunch towards the origin \cite{us1}.

\subsection{Quantum anti--centrifugal force}
\label{twotwo}

We now turn to a seemingly different topic: The Schr\"odinger
equations of a free particle of mass $M$ and energy $E>0$
in two and three dimensions. We recall that in two dimensions
the radial wave equation for vanishing angular momentum ($m=0$)
contains an attractive potential.
This potential is related to
interference of plane waves
in two dimensions as expressed by the Sommerfeld representation
(\ref{somm}) of the zeroth Bessel function. In this way we
make the connection to Huygens' principle.

We start from the time independent Schr\"odinger equation

\begin{displaymath}
-\frac{\hbar^2}{2M}\Delta \Psi(x,y)=E\Psi(x,y)
\end{displaymath}
of the free particle and introduce polar coordinates
$r$ and $\varphi$ by the transformation
$x\equiv r \cos \varphi$ and $y \equiv r \sin \varphi$.
With the help of the {\em ansatz}

\begin{displaymath}
\Psi(x,y)=e^{im\varphi}\frac{u_{m}(r)}{\sqrt{r}}
\end{displaymath}
we arrive at the radial Schr\"odinger equation

\begin{displaymath}
\left\{ \frac{d^2}{dr^2}+\frac{2M}{\hbar^2}
\left[ E-V_{m}^{(2)}(r)\right] \right\} u^{(2)}_{m}(r)=0
\end{displaymath}
Here we have introduced the effective potential \cite{flug}

\begin{displaymath}
V_{m}^{(2)}(r)\equiv \frac{\hbar^2}{2M}
\frac{m^2-1/4}{r^2}
\end{displaymath}
The first term in $V_{m}^{(2)}$,
proportional to $m^2$, is the potential
which describes the familiar centrifugal force.
The negative correction term $-1/4$ gives rise to
a centripetal force
which from this point on we shall call a quantum
anti-centrifugal force to emphasize that its binding
power arises from quantum mechanics.

The effect of this contribution
stands out most clearly for
a particle with zero angular momentum, that is $m=0$.
In this case the effective potential

\begin{equation}
V_{Q}(r) \equiv V_{0}^{(2)}(r)
= -\frac{\hbar^2}{2M}\frac{1}{4r^2}
\label{vq}
\end{equation}
becomes attractive and the radial wave function reads
\begin{displaymath}
u_{0}(r)=\sqrt{r}J_{0}\left( kr \right)
\end{displaymath}
Here and throughout the paper the subscript $0$ on the
wave functions indicates vanishing angular momentum.

Hence, the bunching effect of the zeros of $J_{0}$ takes
on a new meaning. It reflects the fact that there is an
attractive force towards the center $x=y=0$.
Indeed, the force corresponding to $V_{Q}$ causes a
ring--shaped wave packet to spread in asymmetric
way: It initially spreads faster towards the center than
towards the outside \cite{us2}.

In three dimensions no such quantum potential exists.
We recall \cite{bohm} that in this case the effective potential
is proportional to the quantum mechanical square $l(l+1)$
of the angular momentum quantum number $l$. Hence, for
$l=0$ the potential vanishes and no asymmetric spreading
takes place.

\section{Propagators}

This dimensionality--dependent
spreading is a consequence of the propagators being different
in different dimensions. In this section we compare and
contrast the radial propagators in two and three dimensions.
Throughout this section we confine ourselves to radially
symmetric wave functions, that is, wave functions of vanishing
angular momentum. This symmetry is preserved under time
evolution, therefore it suffices to consider the radial
propagator.

We show that the radial propagator of a free
quantum mechanical particle in three dimensions is the
difference between two propagators of the
free one--dimensional radial motion. However, in two dimensions
the propagator is different, reflecting
the attractive quantum anti--centrifugal force. The
mathematical reason for this difference is exactly the same
as in the d'Alembert propagators: It is the difference between a
spherical and an ordinary Bessel function.

\subsection{One dimension}

In this section we lay the foundations for the following sections
by briefly recalling the Green's function $G^{(1)}_{S}$ of the free
Schr\"odinger particle of mass $M$. The Green's function

\begin{equation}
G_{S}^{(1)} \left( x,t\mid \tilde{x},t=0\right)
\equiv {\cal N}(t)\, e^{i\alpha(t)(x-\tilde{x})^2}
\label{g1d}
\end{equation}
with the normalization constant

\begin{equation}
{\cal N}(t)\equiv \sqrt{\frac{\alpha(t)}{i\pi}}
\label{enne}
\end{equation}
and

\begin{equation}
\alpha(t)\equiv \frac{M}{2\hbar t}
\label{alal}
\end{equation}
allows us to propagate the initial wave function
$\Psi^{(1)}_{0}(x,t=0)\equiv \Phi^{(1)}_{0}(x)$
to a later time, such that

\begin{equation}
\Psi^{(1)}(x,t)=\int_{-\infty}^{\infty}\!\! d\tilde{x}\, G^{(1)}_{S}
\left( x,t \mid
\tilde{x}, t=0\right) \Phi^{(1)}_{0}(\tilde{x})
\label{psi1d}
\end{equation}

For the remainder of the article it is useful to recall that for
$t\rightarrow 0$ the parameter $\alpha$ and
the normalization constant ${\cal N}$ approach infinity.

\subsection{Two dimensions}

We now use the expression (\ref{g1d}) for
the propagator $G_{S}^{(1)}$ in one dimension to derive the
Green's function of the radial motion in two dimensions.
For this purpose we take the product of
the one--dimensional Green's functions
of the motions along the $x$- and the $y$-axes together with
the initial wave function
and integrate over the angular and radial variables.
We then express the wave functions by the radial wave
functions.

\subsubsection{Radial propagator}

We start from the generalization

\begin{equation}
\Psi^{(2)}_{0}(x,y;t)=
\int_{-\infty}^{\infty} \!\!\!\!\! d\tilde{x}
\int_{-\infty}^{\infty}\!\!\!\!\! d\tilde{y} \;
G^{(1)}_{S}\! \left( x,t \mid \tilde{x},0\right) G^{(1)}_{S}\!
\left( y,t \mid
\tilde{y},0\right)
\Phi^{(2)}_{0}\!\! \left( \sqrt{\tilde{x}^{2}+\tilde{y}^{2}}\right)
\label{psi2d}
\end{equation}
of the propagation equation
(\ref{psi1d}) to two dimensions. Here we have considered a
rotationally symmetric initial wave function $\Psi^{(2)}_{0}
(\tilde{x},\tilde{y};t=0)\equiv \Phi^{(2)}_{0}(\rho)$,
which depends on the radial variable
$\rho\equiv \sqrt{\tilde{x}^2+\tilde{y}^2}$ only.

When we substitute the expression (\ref{psi1d}) for
the one-dimensional Green's function $G^{(1)}_{S}$
into (\ref{psi2d}) and introduce polar coordinates
$\rho$ and $\varphi$ we arrive at

\begin{eqnarray}
\Psi^{(2)}_{0}(x,y;t)
& = & \Psi^{(2)}_{0}(r;t) \nonumber \\
& = & \int_{0}^{\infty}\!\!\!\!\! d\rho \: {\cal N}
e^{i\alpha (r^{2}+\rho^{2})}
{\cal N}\, \sqrt{\rho}\;\int_{-\pi}^{\pi}\!\!\!\!\! d\varphi
\: e^{-i2\alpha r\rho \cos \varphi} \sqrt{\rho}
\: \Phi^{(2)}_{0} (\rho)
\label{vol2d}
\end{eqnarray}
The wave function $\Psi^{(2)}_{0}(x,y;t)$
depends on the coordinates $x$ and $y$ only through the radial
coordinate $r\equiv \sqrt{x^2+y^2}$. Hence, the axial
symmetry of the initial wave function is preserved under
time evolution.

We recall the Sommerfeld representation (\ref{somm})
of the ordinary Bessel function $J_{0}$, and the
propagation equation reduces to

\begin{equation}
\Psi^{(2)}_{0}(r;t)=\int_{0}^{\infty}\!\!\!\!\! d\rho \: {\cal N}
e^{i\alpha (r^{2}+\rho^{2})}
\frac{1}{\sqrt{i}}\sqrt{\pi \alpha r \rho} \, 2 J_{0}[2\alpha r \rho] \,
\sqrt{\rho}
\: \Phi^{(2)}_{0} (\rho)
\label{volbis}
\end{equation}
where we have made use of the definition
(\ref{enne}) of ${\cal N}(t)$.
With the help of the radial wave functions

\begin{displaymath}
u^{(2)}_{0}(r;t)\equiv \sqrt{r}\: \Psi^{(2)}_{0} \left( r;t\right)
\end{displaymath}
and

\begin{displaymath}
v_{0}^{(2)}(\rho)\equiv u^{(2)}_{0}(\rho;t=0)\equiv \sqrt{\rho}\:
\Phi_{0}^{(2)} \left( \rho \right)
\end{displaymath}
we find

\begin{eqnarray}
u^{(2)}_{0}(r;t) & = & \int_{0}^{\infty}\!\!\!\!\!d\rho \, {\cal N}
\, e^{i\alpha (r^{2}+\rho^{2})} \,
\frac{1}{\sqrt{i}}\sqrt{\pi \alpha r \rho} 2 \, J_{0}[2\alpha r \rho]
\, v_{0}^{(2)} (\rho) \nonumber \\
& \equiv & \int_{0}^{\infty}\!\!\!\!\!d\rho \,
g^{(2)}(r,t\mid \rho) \, v_{0}^{(2)} (\rho)
\label{rad2d}
\end{eqnarray}
This relation suggests the formula

\begin{equation}
g^{(2)}(r,t\mid \rho)\equiv {\cal N}(t) \, e^{i\alpha(t)(r^2+\rho^2)}
\frac{1}{\sqrt{i}}\sqrt{\pi \alpha(t) r \rho}\;
2 J_{0}[2\alpha(t) r\rho]
\label{g22}
\end{equation}
for the Green's function of the radial motion of a free particle in two
dimensions. The key feature of this result is the emergence of the ordinary
Bessel function $J_{0}$.

\subsubsection{Alternative form}

It is instructive to
represent \cite{bia} the Bessel function

\begin{displaymath}
J_{0}(\xi)=\frac{1}{2}\left[ H_{0}^{(1)}(\xi)+
H_{0}^{(2)}(\xi)\right]
\end{displaymath}
in terms of the two Hankel functions

\begin{displaymath}
H_{0}^{(1)}(\xi)=\sqrt{\frac{2}{\pi \xi}}\; e^{i(\xi-\pi/4)}
{\cal H}^{*}(\xi)
\end{displaymath}
and

\begin{displaymath}
H_{0}^{(2)}(\xi)=\sqrt{\frac{2}{\pi \xi}}\; e^{-i(\xi-\pi/4)}
{\cal H}(\xi)
\end{displaymath}
where we have introduced the abbreviation

\begin{equation}
{\cal H}(\xi)\equiv \frac{1}{\sqrt{\pi}}\int_{0}^{\infty}d\zeta
\frac{e^{-\zeta}}{\sqrt{\zeta}\sqrt{1-i\zeta/(2\xi)}}=
\frac{2}{\sqrt{\pi}}\int_{0}^{\infty} d \tau
\frac{e^{-\tau^2}}{\sqrt{1-i\tau^2/(2\xi)}}
\label{scrh}
\end{equation}

We substitute these expressions into the formula (\ref{g22})
for the propagator and complete the squares. We arrive at

\begin{equation}
g^{(2)}= {\cal N} e^{i\alpha (r-\rho)^2}
{\cal H}(2\alpha
r\rho) -i\, {\cal N}e^{i\alpha (r+\rho)^2}{\cal H}^*(2\alpha r\rho)
\label{eq21}
\end{equation}
where we have used the relation

\begin{displaymath}
\frac{1}{\sqrt{i}}=\left( e^{-i\pi/2}\right)^{1/2}=e^{-i\pi/4}
\end{displaymath}

When we compare the radial Green's function $g^{(2)}$ in two dimensions
(\ref{eq21}) to the Green's function $G^{(1)}_{S}$
in one dimension (\ref{g1d}) we find the representation

\begin{equation}
g^{(2)}(r,t\mid \rho ) = G^{(1)}_{S}(r,t\mid \rho){\cal H}[2\alpha(t)
\, r\rho]-
i\, G^{(1)}_{S} (r,t\mid -\rho){\cal H}^*[2\alpha(t)\, r\rho]
\label{g2d}
\end{equation}

Hence the Green's function
$g^{(2)}$ of the radial motion is the difference
of two one-dimensional Green's functions:
One corresponds to the motion
starting at $\rho$, the other at the mirror image $-\rho$.
The two functions reflect the fact that the radial coordinate
is only defined for positive values. In other words, there is an
infinitely steep and infinitely high potential well at $\rho=0$ and the radial
wave function has to vanish there. This fact guarantees that the radial
momentum operator is Hermitian for $r\ge 0$.

Moreover, there is a phase difference of $\pi/2$ between the two
contributions in (\ref{g2d}). This phase difference
is just another manifestation of the
interference of plane waves in two dimensions giving rise
to an ordinary Bessel function as discussed in
section \ref{bhp}.

However, the most important feature of (\ref{g2d}) is the fact that each
contributing term is the product of the Green's function of a
free motion and the function ${\cal H}$. Hence the propagator
of a free particle in two dimensions and of vanishing angular momentum
is not simply that of free motion, but is modified by ${\cal H}$.

\subsubsection{Short time limit}

We can gain deeper insight into the correction factor ${\cal H}$
by considering the short time limit, that is $t\rightarrow 0$.
From section \ref{two} we recall that in this
case $\alpha\rightarrow \infty$. The argument
${\cal H}$ is the product of $\alpha$ and the radial coordinates
$r$ and $\rho$. Hence we are dealing with a rather subtle limit.

For $r \neq 0$ and $\rho \neq 0$ we can consider
the asymptotic limit of ${\cal H}$ for large argument $\xi$.
When we expand the second square root in the definition (\ref{scrh})
of ${\cal H}$ we find

\begin{displaymath}
{\cal H}(\xi) \cong
\frac{2}{\sqrt{\pi}}\int_{0}^{\infty}
d\tau \, e^{-\tau^2}\left( 1+i\frac{\tau^2}{4\xi}\right)
\end{displaymath}
With the help of the integral relations

\begin{displaymath}
\frac{2}{\sqrt{\pi}}\int_{0}^{\infty}
d\tau \, e^{-\tau^2}=1
\end{displaymath}
and

\begin{displaymath}
\frac{2}{\sqrt{\pi}}\int_{0}^{\infty}
d\tau \, e^{-\tau^2}\tau^2=\frac{1}{2}
\end{displaymath}
we arrive at

\begin{displaymath}
{\cal H}(\xi) \cong 1+i\frac{1}{8\xi}
\end{displaymath}
We recall the definition (\ref{alal}) of $\alpha$ and the correction
factor ${\cal H}$ to the free propagator reads

\begin{displaymath}
{\cal H}[2 \alpha(t) r \rho]={\cal H}\left( \frac{M}{\hbar t}r\rho\right)=
1+\frac{i}{\hbar}\frac{\hbar^2}{2M}\frac{1}{4r\rho}t
\end{displaymath}
The second term here is proportional to
the quantum anti--centrifugal
potential $V_{Q}$ defined in (\ref{vq}) and the function ${\cal H}$
takes the compact form

\begin{equation}
{\cal H}[2 \alpha(t) r \rho]=1-\frac{i}{\hbar}
V_{Q}(\sqrt{r\rho})t\simeq \exp\left[ -\frac{i}{\hbar}
V_{Q}(\sqrt{r\rho})t \right]
\label{eq26}
\end{equation}

This formula shows clearly that for short times the radial motion in two
dimensions is determined by the quantum anti--centrifugal
potential $V_{Q}$.

We emphasize that the expression (\ref{eq26}) is only
valid for short times and for radial coordinates $r\neq 0$ and $\rho\neq 0$.
For $t \neq 0$ and $r=\rho=0$ the radial Green's function $g^{(2)}$
vanishes, that is

\begin{displaymath}
g^{(2)}(r=0,t\mid \rho)=g^{(2)}(r,t\mid \rho=0)=0
\end{displaymath}
as required by the boundary condition
$u^{(2)}_{0}(r=0)=0$ and $v^{(2)}_{0}(\rho=0)=0$.
Indeed, this property follows from the exact expression
(\ref{g22}) for $g^{(2)}$ in terms of the Bessel function.

\subsection{Three dimensions}

We now turn to the case of a free particle in three dimensions. Again we
consider a wave function $\Psi^{(3)}_{0}(\tilde{x},
\tilde{y},\tilde{z};t=0)
\equiv \Phi^{(3)}_{0}\left( \sqrt{\tilde{x}^2+\tilde{y}^2+\tilde{z}^2} \right)$
that at $t=0$ is rotationally symmetric, corresponding to vanishing
angular momentum, that is, $l=0$. Time evolution preserves this symmetry.
In order to find the Green's function in three dimensions we
take the product of the Green's functions $G^{(1)}_{S}$ of the free motions
along the $x$--, $y$--, and $z$--axes together with the initial wave function,
and integrate over spherical coordinates $\rho$, $\theta$ and $\varphi$.

We start from the expression

\begin{eqnarray}
\Psi^{(3)}_{0}(x,y,z;t) & = & \int_{0}^{\infty}\!\!\!\!\! d\tilde{x}
\int_{0}^{\infty}\!\!\!\!\! d\tilde{y}
\int_{0}^{\infty}\!\!\!\!\! d\tilde{z} \: \nonumber \\
& & G^{(1)}_{S}(x,t\mid \tilde{x})
G^{(1)}_{S}(y,t\mid \tilde{y})G^{(1)}_{S}(z,t\mid \tilde{z})\Phi^{(3)}_{0}
\left(\sqrt{\tilde{x}^2+\tilde{y}^2+\tilde{z}^2} \right)
\end{eqnarray}
for the propagation of the wave function $\Phi^{(3)}_{0}$.
With the help of the explicit formula (\ref{g1d}) for the Green's
function $G^{(1)}_{S}$ and spherical coordinates we find

\begin{eqnarray}
\Psi^{(3)}_{0}(x,y,z;t) & = & \Psi^{(3)}_{0}(r;t)= \nonumber \\
& = & \int_{0}^{\infty} \!\!\!\!\!
d\rho \, {\cal N} e^{i\alpha (r^2+\rho^2)}{\cal N}^2 \rho
\int_{0}^{\pi} \!\!\!\!\!d\theta \sin \theta \:
e^{-i2\alpha r\rho \cos \theta}
\int_{-\pi}^{\pi}\!\!\!\!\! d\varphi \: \rho \, \Phi^{(3)}_{0} (\rho)
\label{vol3d}
\end{eqnarray}
Since the right hand side depends on the coordinates $x$, $y$, and $z$
only via the radial coordinate $r\equiv \sqrt{x^2+y^2+z^2}$, the
wave function $\Psi^{(3)}_{0}$ at time $t$ only depends on $r$.
Hence, the radial symmetry of the initial wave function is preserved
during time evolution.

The integrand is independent of $\varphi$ and we can immediately perform
this and the $\theta$ integration. Indeed, 
with the help of the integral representation (\ref{sbf})
of the spherical Bessel function $j_{0}$ we arrive at

\begin{displaymath}
\Psi^{(3)}_{0}(r;t)=
\int_{0}^{\infty} \!\!\!\!\!
d\rho \, {\cal N} e^{i\alpha (r^2+\rho^2)} \, 4 \pi {\cal N}^2 \rho
\, j_{0}(2\alpha r \rho) \, \rho \, \Phi^{(3)}_{0} (\rho)
\end{displaymath}
When we introduce the radial wave functions

\begin{displaymath}
u^{(3)}_{0}(r,t)=r\, \Psi^{(3)}_{0}(r,t)
\end{displaymath}
and

\begin{displaymath}
v_{0}^{(3)}(\rho)\equiv u^{(3)}_{0}(\rho,t=0) \equiv
\rho \: \Phi^{(3)}_{0} (\rho)
\end{displaymath}
in three dimensions corresponding to vanishing angular momentum we find
the propagation equation

\begin{eqnarray*}
u^{(3)}_{0}(r,t) & = & \int_{0}^{\infty} \!\!\!\!\!
d\rho \, {\cal N} e^{i\alpha (r^2+\rho^2)} \, 4 \pi {\cal N}^2 \, r \rho
\, j_{0}(2\alpha r \rho) \, v_{0}^{(3)}(\rho) \nonumber \\
& \equiv & 
\int_{0}^{\infty} \!\!\!\!\!
d\rho \, g^{(3)}(r,t\mid \rho ) \, v_{0}^{(3)}(\rho)
\end{eqnarray*}
This approach yields the Green's function

\begin{equation}
g^{(3)}(r,t\mid \rho )\equiv  {\cal N}(t) e^{i\alpha(t) (r^2+\rho^2)}
\, \frac{2}{i} \, 2 \alpha(t) \, r \rho \, j_{0}[2\alpha(t) r \rho]
\label{eq33}
\end{equation}
for the radial motion in three dimensions.
Here we have used the definition
(\ref{enne}) of the normalization constant ${\cal N}$.

When we compare the expression (\ref{eq33}) for the radial propagator in
three dimensions to the corresponding Green's
function (\ref{g22}) in two dimensions we find, apart from some factors,
two changes: (i) In the transition from two to three space dimensions
we have replaced the ordinary Bessel function $J_{0}$ by a spherical
Bessel function $j_{0}$, (ii) moreover, the square root of $i$ in the
denominator in equation (\ref{g22}) is now replaced by $i$.

These two at first
sight minor substitutions have dramatic consequences. As discussed in
equation (\ref{spher}) the spherical Bessel function
is a superposition of an incoming and an outgoing spherical wave with a phase
difference of $\pi$. When we use this representation
of $j_{0}$, we find the Green's function

\begin{displaymath}
g^{(3)}(r,t\mid \rho )\equiv {\cal N}(t)\left[  e^{i\alpha(t) (r-\rho)^2}
-e^{i\alpha(t)(r+\rho)^2}\right]
\end{displaymath}
for the radial motion in three dimensions.
We emphasize that this result is valid for all times.

Similar to the Green's function (\ref{g2d}) in two dimensions, the
corresponding expression in three dimensions is again the difference

\begin{displaymath}
g^{(3)}(r,t\mid \rho)=G^{(1)}_{S}(r,t \mid \rho)-G^{(1)}_{S}(r,t \mid -\rho)
\end{displaymath}
between two propagators of one--dimensional motion:
One corresponds to the motion starting at $\rho$ and the other one at
$-\rho$. However, in contrast to the Green's function in two dimensions,
now we deal with a purely free motion represented by the Green's
function $G^{(1)}_{S}$: The modification factor ${\cal H}$ of two
dimensions is absent since in three dimensions there
is no quantum anti--centrifugal potential.

Moreover, there is a phase shift of $\pi$ between the two contributions.
Its origin is clearly the interference of plane waves in three
dimensions giving rise to a spherical Bessel function.

\section{Conclusions}

We conclude by emphasizing that the radial propagator of a free
particle of vanishing angular momentum moving in three
dimensions is determined by the spherical Bessel function.
It originates from the interference of plane waves in three
dimensions. Its explicit representation in spherical waves shows
that the radial propagator is
the interference of two propagators of free
motion in one dimension. The phase difference between the two contributions
is $\pi$.

In two dimensions the radial propagator is determined by the
ordinary Bessel function. It originates from interference of plane
waves in two dimensions. Since an ordinary Bessel function is not
a superposition of two cylindrical waves but involves correction terms,
the propagator is not a superposition of propagators corresponding to free,
one--dimensional motion. This feature reflects the presence of
the quantum anti--centrifugal force in two dimensions.

In mathematical terms, the dramatic difference between the
radial propagators in two and in three dimensions results from the
different area or volume elements. A comparison between the corresponding
equations (\ref{vol2d}) and (\ref{vol3d}) brings this feature out most
clearly: In two as well as in three dimensions the phases
of the interfering waves, that is the phases
of the exponential in the angular integration
are identical. However, the integration measures are different:
In two dimensions we have $d\varphi$ whereas in three dimensions we have
$d\theta \sin \theta=-d(\cos\theta)$. This difference leads to an ordinary
Bessel function $J_{0}$ in two dimensions but to the spherical Bessel function
in three dimensions.

It is interesting to note that this very same reason also causes
the difference between the Green's functions of the d'Alembert wave equation
in two and three dimensions, and the breakdown of Huygens' principle
in two dimensions.

\section*{Acknowledgements}

We thank M V Berry and I Bia\l ynicki--Birula for fruitful discussions.
Two of us (J P D and M F)
gratefully acknowledge the support of the Humboldt Stiftung
and the great hospitality we have enjoyed at the
Abteilung f\"ur Quantenphysik. D M G was supported in part by the
National Science Foundation Grant PHY97-22614.

\section*{Appendix. Green's function for the d'Alembert wave equation}

In this appendix we briefly derive the Green's function $G^{(N)}$ of the
$N$-dimensional d'Alembert wave equation

\begin{equation}
\left( \frac{1}{c^2}\frac{\partial^2}{\partial t^2}-\Delta^{(N)}\right)
G^{(N)}(t,\vec{r})=\delta(t)\delta(\vec{r})
\label{dal}
\end{equation}
Here $\vec{r}$ and $\Delta^{(N)}$ denote the $N$-dimensional position vector
and Laplacian, respectively.

In order to facilitate the comparison between the Green's function of
the d'Alembert wave equation and
the Schr\"odinger equation of a free particle in
the main body of the paper we focus on the cases of
two and three dimensions. We show that in three dimensions the Green's
function of the d'Alembert equation
is a delta function. In contrast, in two dimensions the Green's function
carries a tail behind a sharp edge. We trace this difference back to
the difference between the spherical and the ordinary Bessel functions.

\subsection*{Fourier solution in $N$ dimensions}

In order to solve the inhomogeneous wave equation (\ref{dal})
we make the Fourier {\em ansatz}

\begin{displaymath}
G^{(N)}(t,\vec{r})=\int d k_{0}
\int d^{N}k \, \tilde{G}(k_{0},\vec{k})\,
e^{i(k_{0}ct-\vec{k}\cdot \vec{r})}
\end{displaymath}
where $\vec{k}$ denotes the $N$-dimensional wave vector.

When we substitute
this {\em ansatz} into (\ref{dal}) and use the Fourier representation

\begin{displaymath}
\delta(t)\delta(\vec{r})=\frac{1}{(2\pi)^{N+1}}
\int d k_{0} \int d^{N}k \, e^{ik_{0}c t}
e^{-i\vec{k}\cdot \vec{r}}
\end{displaymath}
of the product of delta functions, we find the algebraic equation

\begin{displaymath}
\left[ k_{0}^2 - \vec{k}^2 \right]
\tilde{G}(k_{0}, \vec{k} )=-\frac{1}{(2\pi)^{N+1}}
\end{displaymath}
with the solution

\begin{displaymath}
\tilde{G}(k_{0} , \vec{k} )=-\frac{1}{(2\pi)^{N+1}}
\frac{1}{k_{0}^2-\vec{k}^2}=
-\frac{1}{(2\pi)^{N+1}}\frac{1}{2 k}
\left[ \frac{1}{k_{0}-k}
-\frac{1}{k_{0}+k} \right]
\end{displaymath}
where $k\equiv \mid \vec{k} \mid$.

Hence, the Fourier transform $\tilde{G}$ enjoys simple
poles at $k_{0}= \pm k$. Different paths in the complex plane
$k_{0}$ define different Green's functions.
Since in the present paper we are only interested in the dependence
of the Green's function on the
dimensions we choose one particular propagator,
namely the retarded propagator $G^{(N)}_{\rm{ret}}$.

The retarded Green's function is defined by a
path just below the real $k_{0}$-axis. For $t<0$
we can close the path by a circle at infinity in the
lower complex plane without changing the value
of the integral. Since in this case we do not
include any poles we find

\begin{displaymath}
G^{(N)}_{\rm{ret}}(t,\vec{r})=0
\end{displaymath}
On the other hand when $t>0$
we can close the path in the upper complex
plane without changing the value of the integral.
The residue theorem

\begin{displaymath}
\oint dz \frac{f(z)}{z-z_{0}}=2\pi i f(z_{0})
\end{displaymath}
where the integration path circumvents the pole $z_{0}$
in the counter--clockwise direction allows us to perform the
integration over $k_{0}$.
We find the expression

\begin{displaymath}
G^{(N)}_{\rm{ret}}=\frac{(-i)}{(2\pi)^N}\int \frac{d^N k}{2k}
\left( e^{ikct}
-e^{-ikct}\right) e^{-i\vec{k}\cdot \vec{r}}
\end{displaymath}
We can combine the results for positive and negative values of $t$
in the form

\begin{equation}
G^{(N)}_{\rm{ret}}(t,\vec{r})= \Theta(t)
\frac{1}{(2\pi)^N}\int d^N k\frac{\sin (kct)}{k}
e^{-i\vec{k}\cdot \vec{r}}
\label{bistar}
\end{equation}

It is instructive to cast this formula into a slightly different
form by separating the $N$--dimensional volume element \cite{sommer}

\begin{displaymath}
d^{N}k=k^{N-1}dk \, d^{N-1}\omega
\end{displaymath}
into the integration over the length of the $N$--dimensional wave vector
and the $N-1$ angles contained in $d\omega$.

The Green's function then reads

\begin{displaymath}
G^{(N)}_{ret}(t,\vec{r})=\Theta(t)\frac{1}{(2\pi)^N}
\int_{0}^{\infty} dk k^{N-2}\sin (kct) \int d^{N-1}\omega e^{-i\vec{k}
\cdot \vec{r}}
\end{displaymath}
The retarded Green's function of the $N$--dimensional d'Alembert wave
equation is determined by a two--fold interference: (i) the interference
of plane waves in the $N-1$ dimensional space of angles and (ii) the
interference of these expressions due to the integration over all wave
numbers. In (i) all plane waves have the same amplitude and the same
wave number. However, they differ in their propagation direction.
All directions appear with equal weight but the surface element
can put different weight on different directions. It is this
contribution that creates the difference between the propagator in two
and in three dimensions, as we show now.

\subsection*{Two dimensions}

We first consider the case of two dimensions, that is $N=2$ and perform the
remaining integrations over the two--dimensional wave vector $\vec{k}$
in polar coordinates. With the help of the area element

\begin{displaymath}
d^2 k=dk\:k \: d\varphi
\end{displaymath}
and the Sommerfeld representation (\ref{somm}) of the Bessel
function $J_{0}$, the Green's function

\begin{displaymath}
G^{(2)}_{\rm{ret}}= \Theta(t)\frac{1}{2\pi} \int_{0}^{\infty} dk
\sin (kct)\frac{1}{2\pi}
\int_{-\pi}^{\pi} d\varphi \, e^{-ikr\cos \varphi}
\end{displaymath}
following from (\ref{bistar}) takes the form

\begin{equation}
G^{(2)}_{\rm{ret}}= \Theta(t)
\frac{1}{2\pi}\int_{0}^{\infty} dk \sin (kct) J_{0}(kr)
\label{star}
\end{equation}
The integral relation \cite{grad}

\begin{displaymath}
\int_{0}^{\infty}d\xi \sin
(\xi a)J_{0}(\xi b)=\frac{\Theta(a^2-b^2)}{\sqrt{a^2-b^2}}
\end{displaymath}
finally yields the expression

\begin{equation}
G^{(2)}_{\rm{ret}}(t,\vec{r})\equiv
G^{(2)}_{\rm{ret}}(t,r)=\Theta(t)
\frac{1}{2\pi}\frac{\Theta[(ct)^2-r^2]}{\sqrt{(ct)^2-r^2}}
\label{app1}
\end{equation}
for the retarded Green's function of the d'Alembert wave
equation in two dimensions.

This Green's function enjoys a sharp wave front due to
the Heaviside step function. Moreover, it is followed by a decay
which obeys an inverse square root dependence.

\subsection*{Three dimensions}

We now turn to the three--dimensional case where the volume element in
wave vector space reads

\begin{displaymath}
d^3k=dk\; k^2 d\theta \sin \theta \: d \varphi
\end{displaymath}
The retarded Green's function then takes the form

\begin{displaymath}
G^{(3)}_{\rm{ret}}=\Theta(t)
\frac{1}{(2\pi)^2}\int_{0}^{\infty}dk \sin (kct)
k \int_{0}^{\pi}d\theta \sin \theta e^{-ikr\cos
\theta}\frac{1}{2\pi}\int_{-\pi}^{\pi}d\varphi
\end{displaymath}
which after simple angular integrations with
the help of the integral
representation (\ref{sbf}) of the spherical Bessel function $j_{0}$
reduces to

\begin{displaymath}
G^{(3)}_{\rm{ret}}
=\Theta(t) \frac{1}{\pi}\frac{1}{2\pi}\int_{0}^{\infty}
dk \sin (kct) \, k \, j_{0}(kr)
\end{displaymath}
When we compare this expression to the corresponding formula
(\ref{star}) for two dimensions we recognize that the ordinary
Bessel function $J_{0}$ has been replaced in three dimensions by
the spherical Bessel functions $j_{0}$. This has important consequences
for the remaining integration over $k$.
Indeed, when we use the representation (\ref{sbf}) of
the spherical Bessel function as a sine function we find

\begin{displaymath}
G^{(3)}_{\rm{ret}}
=\Theta(t) \frac{1}{\pi r}\frac{1}{2\pi}\int_{0}^{\infty}
dk \sin (kct) \, \sin (kr) 
\end{displaymath}
which with the help of the trigonometric relation

\begin{displaymath}
\sin \alpha \sin \beta =\frac{1}{2}\left[ \cos (\alpha -\beta)
-\cos (\alpha +\beta) \right]
\end{displaymath}
yields

\begin{displaymath}
G^{(3)}_{\rm{ret}}=\Theta(t)
\frac{1}{2\pi r}\frac{1}{2\pi}\int_{0}^{\infty}
dk \{ \cos [k(ct-r)]-\cos [k(ct+r)]\}
\end{displaymath}
or

\begin{displaymath}
G^{(3)}_{\rm{ret}}=\Theta(t)\frac{1}{4\pi r}
[\delta(ct-r)-\delta(ct+r)]
\end{displaymath}
Due to the $\Theta$ function, the retarded Green's function is only
non--vanishing for $t>0$. Hence,
only the first delta function makes a contribution
and we arrive at the expression

\begin{displaymath}
G^{(3)}_{\rm{ret}}=\Theta(t)\frac{\delta(ct-r)}{4\pi r}
\end{displaymath}
for the retarded Green's function of the d'Alembert
wave equation in three dimensions. The integration of
the spherical Bessel function instead of the ordinary
Bessel function has created the delta function in the propagator
instead of the square root and $\Theta$ function.

\end{document}